\documentclass{article}
\usepackage[margin=1.20in]{geometry}

\usepackage{graphicx}
\usepackage{subfigure}
\usepackage{amssymb}
\usepackage{amsmath}

\usepackage[square,numbers]{natbib}

\usepackage{hyperref}
\hypersetup{
    colorlinks=true,
    linkcolor=black,
    citecolor=black,
    filecolor=black,
    urlcolor=black,
    linkbordercolor = {1 1 1},
}

\newcommand{\footremember}[2]{%
   \footnote{#2}
    \newcounter{#1}
    \setcounter{#1}{\value{footnote}}%
}
\newcommand{\footrecall}[1]{%
    \footnotemark[\value{#1}]%
} 

\title{Simulating and Evaluating Rebalancing Strategies \\ for Dockless Bike-Sharing Systems}

\author{Damian Barabonkov\footremember{2}{Authors contributed equally.},
Samantha D\textsc{\char13}Alonzo\footrecall{2},
Joseph Pierre\footrecall{2}
\\ Massachusetts Institute of Technology \\ \texttt{\{damianb,sdalonzo,jopierre\}@mit.edu} 
\\[2ex] 
Daniel Kondor, Xiaohu Zhang\footnote{Data collection and partial preprocessing.}, Mai Anh Tien \\ Singapore-MIT Alliance for Research and Technology \\ \texttt{\{daniel.kondor,xiaohu,mai.tien\}@smart.mit.edu}
}
\date{}

\begin{document}

\maketitle

\section*{Abstract}

\noindent Following the growth of dock-based bike sharing systems as an eco-friendly solution for transportation in urban areas, \textit{Dockless} systems are revolutionizing the market for the increased flexibility they offer to users. Bike redistribution is a common approach to improve service, and there exists extensive research considering static and dynamic rebalancing strategies for dock-based systems. We approach the dockless problem by defining abstract stations from trip start and end location frequency. This paper offers a optimizing \textit{Mixed Integer Program} framework to model the effects of various bike repositioning strategies for dockless systems. We process 30 days worth of Singapore-based dockless bike data from September 2017 to extract trips. Pairing our mixed integer program with a demand model built from the processed data, we unveil trends between fleet size, lost demand, and magnitude of repositioning proper to the repositioning strategy employed. We also show that increasing repositioning potential does not always improve service performance.

\noindent \emph{Keywords}: Bike Sharing $|$ Dockless $|$ Rebalancing $|$ Optimization

\section{Introduction}
\label{intro}
In recent years, Bike Sharing Systems (BSS) have become more prevalent in major cities in response to global climate change and inconveniences from heavy traffic congestion \cite{Understanding}. BSS are bicycle programs intended to provide an alternative to private transportation and improve first and last mile connections to other modes of transit. While BSS were introduced in Amsterdam in the 1960s, they did not become globally popular until the early 2010s after technological advances allowed for more efficient payment methods and more accurate bike tracking systems \cite{citylab}. Today, BSS exist in over 1,000 cities with a global fleet of nearly 1.5 million shared bicycles \cite{Ghosh}. 

Currently, there are two main types of BSS: dock-based and dockless. In a dock-based BSS, users take bikes from and return them to fixed docks or stations strategically placed around a city. Dock-based BSS and associated problems related to imbalance between supply and demand of bikes at stations have been studied in depth. Ghosh et al presents a thorough summary of existing literature on dock-based BSS and repositioning techniques \cite{Ghosh}. In a dockless BSS bikes are picked up and dropped off from any location within the city that allows bike parking. Dockless bikes are located, unlocked, and paid for through a smart phone app. The focus of the current paper is performing repositioning efficiently in a dockless system using data from Singapore as a test case.

Repositioning or rebalancing is a strategy used by both dock-based and dockless system operators in an attempt to mitigate lost demand. It involves moving bikes from congested locations (too many bikes) to starved stations (too few bikes). Repositioning can be dynamic, which involves moving bikes throughout the day to fix imbalances during peak hours, or static, which involves resetting the system to predetermined initial conditions overnight when demand is negligible \cite{Understanding}. 

Even with rebalancing, imbalances between supply and demand caused by limited bike capacity at each station exist in dock-based BSS. In addition, dock-based systems are limited by the ease of access to docking stations \cite{Understanding}. Dockless systems do not face these issues.  

As a result, Dockless BSS have increased in popularity throughout Asia, particularly in China where two free-floating bike-sharing companies dominate the market \cite{sustainable}. After populating China with millions of bikes, these two Chinese companies launched dockless campaigns in Singapore in early 2017. During the same time, multiple Singaporean companies also launched dockless BSS in Singapore. These companies have dealt with many issues as the dockless bike-sharing movement in Singapore has struggled to gain momentum \cite{Understanding}. There problems were problems from the very beginning, mainly because three separate companies flooded the Singaporean market in 2017. During this time, the fleet size of bikes grew to be unmanageable and utilization rates of bikes were extremely low. Singapore's Land Transit Authority (LTA) was forced to take restrictive measures on dockless BSS via an Operator License in 2017. This restrictive license, combined with low utilization rates leading to low revenue, led to the decline of dockless BSS in Singapore as multiple bike-sharing companies exited the market.   

The issues faced in Singapore's dockless BSS are not unique. Other dockless systems around the world, including shared bikes and shared electric scooters, have been facing similar problems in recent years. These problems, such as mismanagement of dockless fleets and low revenue for companies, may be avoidable in the future with the benefit of proper research and optimization models for dockless systems. When re-positioned and managed properly, BSS provide an environmentally friendly and congestion reducing alternative to private transportation in cities, as well as promote a healthy lifestyle among citizens. Specifically, dockless sharing systems provide users with more flexibility and ease of access than dock-based systems. In this paper, we propose a new method for modeling and optimizing repositioning tasks of dockless BSS based off the work of Ghosh et al \cite{Incentivizing}.  

We propose a data-driven methodology to define abstract stations and extract trips in dockless BSS. We then develop a mixed-integer program (MIP) to generate optimal repositioning tasks for dockless BSS. We demonstrate the applicability of our model by studying the dockless BSS in Singapore during September 2017, when the aforementioned companies flooded the market. Since little research has been done on repositioning methods in dockless BSS, we use our MIP to evaluate lost demand, calculated via a demand model based on historical trip data, in the presence of two different repositioning strategies: a modified form of static repositioning and dynamic repositioning.  We focus on a subset of the data from the Northern region of Singapore for computational efficiency. Since the LTA Operator License expected dockless BSS companies to take a more conservative approach to fleet size and exhibit higher bike utilization rates in the future \cite{transportguru}, a main focus of our research is exploring the trade-offs between repositioning strategies and lost demand in the presence of a diminishing fleet size. We believe our methodology can also be easily adapted for other shared mobility systems, such as electric scooters, and easily modified to study different objectives, such as environmental impact of such systems. 

The remainder of the paper is organized as follows. Section \ref{models} discusses the demand model developed to estimate the demand for bikes at each station in the dockless BSS and the mixed integer program (MIP) developed to optimize the dockless BSS in the presence of different repositioning strategies. Section \ref{definitions} proposes the definition for trips and stations which were used in this research and can be applied to the general case of a dockless BSS. Section \ref{results} presents our results and and Section \ref{future} discusses future research directions. 

\section{Models}
\label{models}
We develop two models in our research. First, we develop a demand model based on Poisson random variables to generate demand scenarios. We then feed the demand scenarios into our second model: a mixed integer program (MIP) that minimizes lost profit through repositioning. 

\subsection{Demand Model}
\label{demandmodel}
The demand model works by polling a Poisson process generated from historical bike trip data. The data consists of an origin station, destination station and timestamp for every trip. However in dockless BSS, there is no notion of stations, and the raw data consists of a collection of location pings with bike IDs which must get processed and clustered into abstract stations. This whole procedure is described later in Section \ref{definitions}. Once processed, we combine data from certain days of the week that exhibit similar ridership timings and patterns in order to increase the amount of data at hand when generating the Poisson process. Data from Monday through Thursday forms one category, Friday another, and Saturday and Sunday a third. These groupings are based on an analysis of variance across weekdays within our data set and findings presented by Xu et al \cite{unraveling}. Then following a similar data aggregation strategy as \cite{Markov}, the bike trip data is sorted into hourly bins per each category based on the start time and weekday of the trip. Then for each hour for every station, a rate of bikes out (demand) on 2 minute intervals is recorded to form a Poisson process. To generate the demand simulation, for every station, the demand random variable is polled as a Poisson random variable 30 times due to the 2 minute intervals and the sum of the bikes out is the demand at that station for that hour. To generate complete bike trips with assigned origins and destinations, a probability distribution per station is built per hour with 2-minute polling intervals for the destinations of trips from historical bike data. Then for every station and its generated demand-out, the respective destination random variable is polled the demand-out number of times to assign destinations and create complete trips. Furthermore, since the trips have assigned destinations, this creates an implicit model for supply when counting the number of trips that have a given station set as their destination. Using this technique, the demand simulation can generate trips for every station at every hour of the week.

\subsection{Mixed Integer Program}
\label{MIP}
In this section we describe the generic model of the Dockless Repositioning Problem. We modify the objective function, inputs, and a few constraints from the Mixed Integer Linear Program (MILP) proposed by Ghosh et al \cite{Incentivizing} to develop a Mixed Integer Program (MIP) to represent the dockless case. The problem can be represented using the following tuple:

\begin{equation*}
    \langle \mathcal{S}, \mathcal{V}, \mathcal{C^*}, d^{t}, \mathcal{F}^{t}, \mathcal{N} \rangle.
\end{equation*}

$\mathcal{S}$ denotes the set of abstracted base stations and $d^{t}_s$ represents the number of bikes at station $s \in \mathcal{S}$ at time-step $t$. We have a set of repositioning vehicles where $C_{\textit{v}}^{*}$ represents the number of bike slots in the vehicle $v \in \mathcal{V}$. The $\mathcal{F}^{t}$ represents a set of $\mathcal{K}$ discrete training demand scenarios generated by the aforementioned demand model at time-step $t$. More concretely, $F_{s,\textit{k}}^{t+}$  and $F_{s,\textit{k}}^{t-}$ represent the number of bikes entering and exiting station $s$ in demand scenario $k$ at time $t$ respectively . Lastly, $N_p$ is the maximum number of pickup locations at each time-step for each vehicle and $N_d$ is the maximum number of drop off locations at each time-step for each vehicle. 

Solving this MIP returns an optimal repositioning strategy for only a single time-step $t$. In order to optimize over a time range, the LP must be solved for increasing values of $t$ and station bike amounts updated after each time-step. This iterative process is described further in Section \ref{MIPiteration}.

For simplicity purposes, we make the following assumptions. First, we assume that a repositioning vehicle can make only one repositioning trip per time-step by setting $N_p$ and $N_d$ to 1. This includes time spent collecting bikes within the neighborhood of the abstract initial station and placing the bikes within the neighborhood of the abstract final station. We also assume all bikes within the confines of the abstract station are accessible and homogeneous, meaning they are all equally likely to be selected by users for a trip or selected by the operator for a repositioning task. 

To represent the repositioning tasks, we introduce the following decision variables:

\begin{itemize}
  \item $y_{s, v}^+$ denotes the number of bikes dropped off by vehicle $v$ at station $s$;
  \item $y_{s, v}^-$ denotes the number of bikes picked up by vehicle $v$ from station $s$;
  \item $b_{s, v}^+$ is a binary decision variable. It is set to 1 if vehicle $v$ dropped off any bikes at station $s$ and 0 otherwise.
  \item $b_{s, v}^-$ is a binary decision variable. It is set to 1 if vehicle $v$ picked up any bikes from station $s$ and 0 otherwise.
  \item $L_s^k$ denotes the lost demand (as a negative number) at station $s$ for demand scenario $k$.
\end{itemize}

We can now describe the method for computing repositioning tasks for the vehicles.  

\begin{align}
&\text{minimize}& &\displaystyle ||\mathcal{K}|| \cdot \alpha \cdot \sum_{s,v}{\min{(y_{s,v}^+, 1})} - \beta \cdot \sum_{s,k}{L_s^k}& \quad\\
&\text{subject to}& &L_s^k \leq d_s^{ t} + \displaystyle \sum_t^T (F_{s,k}^{t+} - F_{s,k}^{t-}) + \displaystyle \sum_{\textit{v}} (y_{s, \textit{v}}^+ - y_{s, \textit{v}}^-)& &\forall{k,s}\\
&& & y_{s, v}^- \leq b_{s,v}^- \cdot d_s^t & &\forall{s,v}\\
&& & \displaystyle \sum_{s} y_{s, v}^- \leq C_v^*& &\forall v\\
&& & \displaystyle\sum_{v} y_{s, v}^- \leq d_s^{ t}& &\forall s\\
&& & y_{s, v}^+ \leq b_{s,v}^+ \cdot C_v^*& &\forall s,v\\ 
&& & \displaystyle\sum_{s} y_{s, v}^+ = \displaystyle\sum_{s^{\prime}} y_{s^{\prime}, v}^-& &\forall v\\
&& & \displaystyle \sum_{s} b_{s,v}^- = N_p& &\forall v\\
&& & \displaystyle \sum_{s} b_{s,v}^+ = N_d& &\forall v\\
&& & b_{s,v}^-, b_{s,v}^+ \in \{0, 1\}; 0 \leq y_{s, v}^+, y_{s, v}^- \leq C_v^*; L_s^k \leq 0
\end{align}

\begin{itemize}
    \item \textbf{Objective (1): Minimize lost profit.} Lost profit is computed as the difference between the number of repositioning trips to handle all $\mathcal{K}$ demand scenarios for all $v \in \mathcal{V}$ and all $s \in \mathcal{S}$ multiplied by some cost parameter $\alpha$ represented by the first term, and the sum of lost demand for each $s \in \mathcal{S}$ during each $k \in \mathcal{K}$ times some cost parameter $\beta$ represented by the second term. These cost parameters provide a simple way of comparing performance given an estimated cost and varying business models (whether the company wishes to prioritize revenue or service quality). The purpose for optimizing over the $\mathcal{K}$ demand scenarios is to come up with a single repositioning combination that works well in the general case and is not too over-fit. The program's iterating frequency equates to rebalancing frequency and can be adjusted as well. Decisions are made one time-step at time and the program is iterative.  
    \item \textbf{Constraint (2): The lost demand is the deficit between the net of the demand at a station and the bikes available at that station.} The absolute value of $L_s^k$ is the number of bike trips that were lost due to improper positioning of bikes in the system. This value is the sum  of the number of bikes present at $s$, the net flow of bikes computed via the demand model for $T$ time-steps into the future and the net change in bikes due to repositioning. The adjustable parameter $T$ can be used to account for the differences in static and dynamic repositioning strategies. 
    \item \textbf{Constraint (3): A vehicle can not pick up more bikes than the amount present at the station.} For a vehicle $v$ and a station $s$, the number of bikes picked up at $s$ at time-step $t$ may not exceed the amount present there at time-step $t$. This also ensures that a vehicle can only pick up bikes from a station it visits ($b_{s,v}^-$ is set to 1). 
    \item \textbf{Constraint (4): A vehicle can not pick up more bikes than its capacity.} For a vehicle $v$, the sum of all of the bikes it picks up may not exceed its capacity.
    \item \textbf{Constraint (5): The number of bikes picked up among all vehicles may not exceed the amount present at the station.} The sum of the bikes picked up at $s$ at time-step $t$ over all vehicles $v$ must be less than or equal to the number of bikes present at $s$ at time-step $t$. 
    \item \textbf{Constraint (6): A vehicle cannot drop off more bikes than its capacity.} The number of bikes dropped off at station $s$ by a vehicle $v$ must be no larger than the capacity of that vehicle $v$. This also ensures that a vehicle can only drop off bikes at a station it visits ($b_{s,v}^+$ is set to 1). 
    \item \textbf{Constraint (7): A vehicle must drop off the same number of bikes it picked up.}
    \item \textbf{Constraint (8): A vehicle can only drop off at a fixed number of stations during a time-step.} During dynamic repositioning, a vehicle can only visit one station in a time-step. During static repositioning, a vehicle can visit up to $N_d$ stations.
    \item \textbf{Constraint (9): A vehicle can only pick up from a fixed number of stations during a time-step.} During dynamic repositioning, a vehicle can only visit one station in a time-step. During static repositioning, a vehicle can visit up to $N_p$ stations. 
\end{itemize}

A mathematical caveat allows the MIP to have a binary decision variable $b$ to be set to 1 while still having its respective $y$ at 0. This explains why $N_p$ and $N_d$ refer to an upper-bound on the number of stations as a repositioning trip is only scheduled if the $y$ is non-zero. This reason is also why the objective function looks at the $y_{s,v}^+$ when counting the number of drop-off trips rather than $b_{s,v}^+$ directly.

Since BSS providers may develop different business models depending on their environment and priorities, we design the MIP to be as flexibile as possible. Parameters such as cost, number of repositioning vehicles, rebalancing frequency, etc. can be easily modified to fit different scenarios and systems. 

This MIP is not formulated as a traditional linear program. However, we were able to program it directly using the Python extension of the IBM ILOP CPLEX Optimization Studio version 12.9. 

\subsection{MIP Iteration}
\label{MIPiteration}

The MIP described solves only for the optimal repositioning strategy at a single time-step $t$ while considering $T$ time-steps of demand into the future. In order to optimize over a time range, the MIP must be iterated over increasing $t$ time-steps and station bike amounts updated with time. When iterating over the MIP to extract results, we choose to use a time-step $t$ of 1 hour which aligns itself well with the demand model setup. Additionally since the MIP considers $T$ hours into the future, as a design decision, we choose to rebalance only once every $T$ hours so to not rebalance on an artificially inflated demand. This effectively treats the $T$ as the rebalancing frequency as well. So upon iterating, on time-steps where the MIP should not rebalance, setting $||\mathcal{V}|| = 0$ effectively turns off rebalancing, and then restoring the $\mathcal{V}$ re-enables the rebalancing.

We choose the genesis $t = 0$ to be on a Monday at midnight. At $t = 0$, for every $s \in \mathcal{S}$, its $d_s^0$ is set to a respective initial amount of bikes calculated as per Section \ref{station}. The demand model is polled for all $\mathcal{F}^{t}$ considering $T$ hours into the future, and then the MIP is populated with all of the parameters of the respective rebalancing strategy and solved. Once solved, the repositioning combinations $y_{s,v}^+$ and $y_{s,v}^-$ are recorded. Then a random $k_R$ is chosen from the $\mathcal{K}$ to be used as the guiding demand scenario at time-step $t$ when gathering results. The $\sum\limits_{s}{L_s^{k_R}}$ is recorded as the lost demand at time-step $t$ and, before the next iteration, the station bike amounts are updated as $d_s^{t+1} \leftarrow \max \Big\{d_s^t + (F_{s,k_R}^{t+} - F_{s,k_R}^{t-}) + \sum\limits_{\textit{v}} (y_{s, \textit{v}}^+ - y_{s, \textit{v}}^-), 0 \Big\}$. In both the static and dynamic repositioning strategies, the MIP is for the same number of iterations to facilitate a fair comparison.

\section{Data Preprocessing}
\label{definitions}

For portability across data-sets, we define the MIP to take trip data as input. However, our data set for dockless BSS in Singapore is a collection of GPS pings rather than defined trips. Our data-set is entries consisting of a bike ID number, a time-stamp, and associated GPS coordinates reported to a central server by the embedded GPS sensor and communication module in each bike. The data ranges over a one month period, from the 1$^{\text{st}}$ to 30$^{\text{th}}$ of September, 2017 and contains 47,302 unique bike IDs. In order to extract trips from the GPS pings in our data-set, we pre-process it as discussed in Section \ref{trips}. Section \ref{station} expands on our approach and simplification of the dockless problem by defining abstract stations, and Section \ref{usage} discusses bike usage trends observed through our Singaporean data-set.

\subsection{Trip Definition}
\label{trips}
Inspired from Shen et al. (2018) \cite{Understanding}, we filter through the data for each bike by making three basic assumptions about trips. First, the time between the consecutive pings should be at least 180 seconds; we assume that users would not take a bike for less than 3 minutes. Second, the GPS locations between two time-steps must change by at least 200 meters for the change in location to be considered a trip; this filters out inconsistencies due to instability in GPS recordings. Finally, the velocity of the trip should be lower than 25 km/hour (the speed limit for bikes in Singapore); this check accounts for potential “trips” due to repositioning, where the bike is in a motored vehicle. In the event that all these conditions are satisfied, a new trip is defined.

\subsection{Station Definition}
\label{station}

\begin{figure*}[h]
\centering
\subfigure[Singapore Clustered]{
 \includegraphics[width=.45\linewidth]{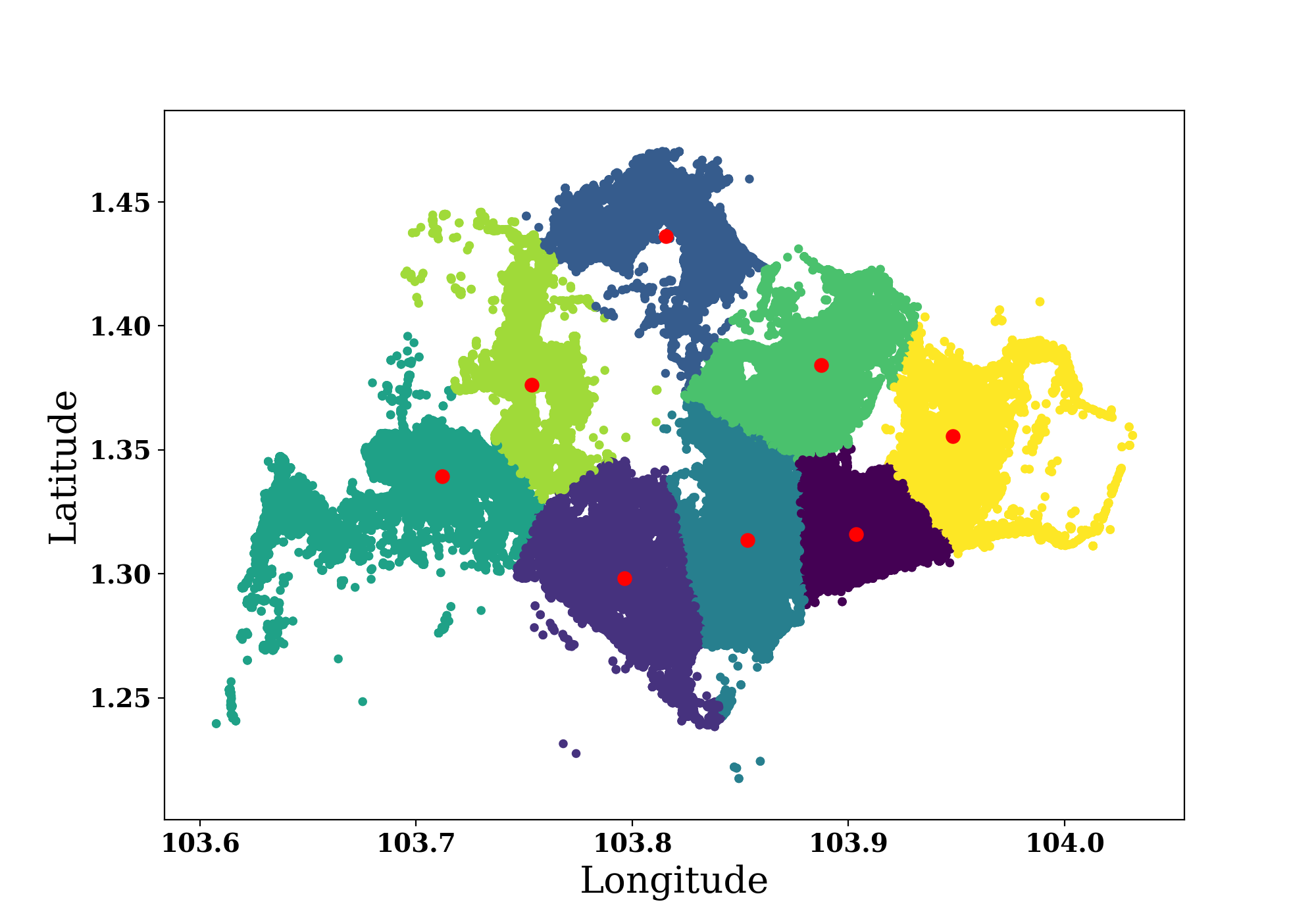}}
\qquad
\subfigure[Woodlands Clustered]{
  \centering
  \includegraphics[width=.45\linewidth]{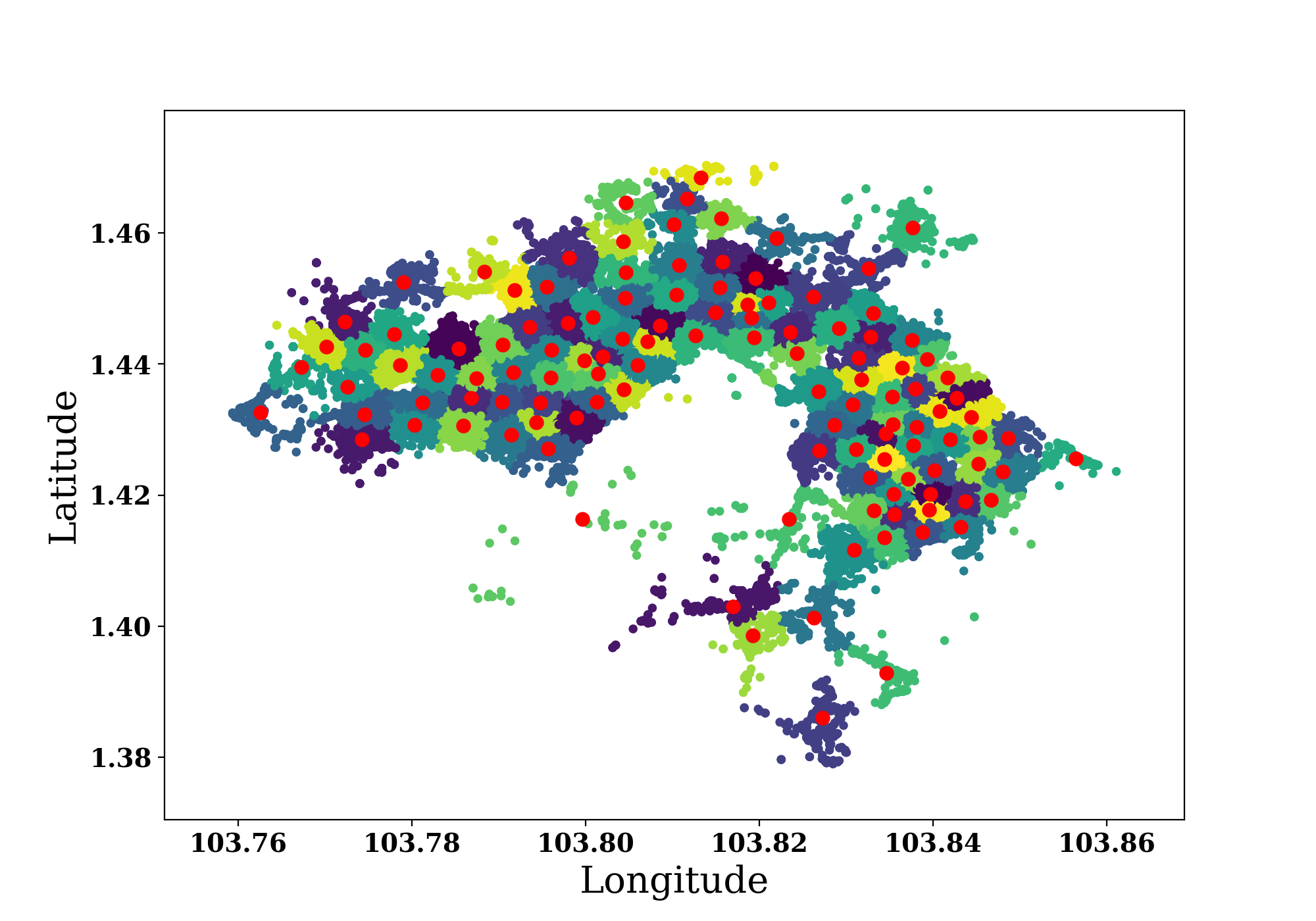}}
\caption{The results of k-means clustering on Singapore (a) and the Woodlands region (b). }
\label{clustered}
\end{figure*}

In order to simplify the modeling, Singapore was segmented into eight separate regions based on the trip data described in Section \ref{trips}. The origin and destination of all 939,762 trips were considered as individual points. The resulting 1,879,524 points were then clustered into eight regions using k-means clustering. 

We focused specifically on the k-means defined Northern region of Singapore labeled in Figure \ref{clustered}. This region roughly correlates to Singapore's Woodlands, a residential town which was selected as a microcosm of Singapore \cite{woodlands}.  We refer to this area as the Woodlands for the remainder of the paper.

The selected region was then segmented again using a k-means clustering approach with 120 clusters. Trips that started in the Woodlands but ended in a different region or vice versa comprised less than 1\% of the roughly 120,000 trips and were filtered out for simplicity. In addition, of the 7,821 bikes in this region, 890 bikes were idle throughout the entire observation period for unknown reasons, leaving 6,931 active bikes. The idle bikes were not considered when constructing the abstract stations as they had no associated trips.

The resulting 120 clusters are treated as abstract stations. To understand the feasibility of and develop a physical representation for each station, a radius and an area were estimated for each station. The radius is determined by calculating the Euclidean distance between the generated k-means central coordinates of the station and the furthest origin or destination point within that cluster. From the radius, the circular area of each station was also computed. 

Each abstract station can be physically represented by the central coordinates and all of the bikes that lie within the computed radius of that coordinate. Of the 120 stations in the Woodlands, 75\% of the stations are less than 1 square kilometer in area. All but seven stations have a radius less than 1 kilometer. The mean radius of the 120 stations is 0.50 km, and the mean area of the 120 stations is 1.01 $\text{km}^2$. This radius and area motivated the selection of 120 as the number of stations because it simplified the dockless problem through abstraction while still producing stations that covered a small enough land area to be feasibly traversed by a dockless BSS operator for repositioning. The stations were granular enough such that only 13$\%$  of trips start and end at the same abstract station. More stations may be added to increase accuracy given enough computational power to iterate the MIP.  

To estimate the initial number of bikes at each station, we mapped the first GPS ping for each of the bikes present during the period of data collection. We then assigned each bike to the closest station based on Euclidean distance.  

Visual representations of the 8 regions of Singapore and the 120 abstract stations in the Woodlands are shown in Figure \ref{clustered}. 

\subsection{Bike Usage}
\label{usage}

As previously mentioned, unmanageable fleet size was a large reason for the LTA's conservative regulations in Singapore's dockless BSS. After being launched in 2017, dockless bikes were parked in inconvenient places, clogging up sidewalks, as well as MRT exits and entrances. Bike maintenance was difficult and faulty bikes sat on the streets for weeks or even months \cite{Understanding}. As a result, the Land Transport Authority (LTA) took serious restrictive measures in 2018 to regulate Dockless BSS. Two major companies had to surrender their operating licenses in Singapore and many other dockless BSS companies withdrew from the Singaporean market. In order to assure that our model and data set align with these facts, we investigate trends in bike usage in our data set. 

\begin{figure}[h]
\centering
\includegraphics[width=.7\linewidth]{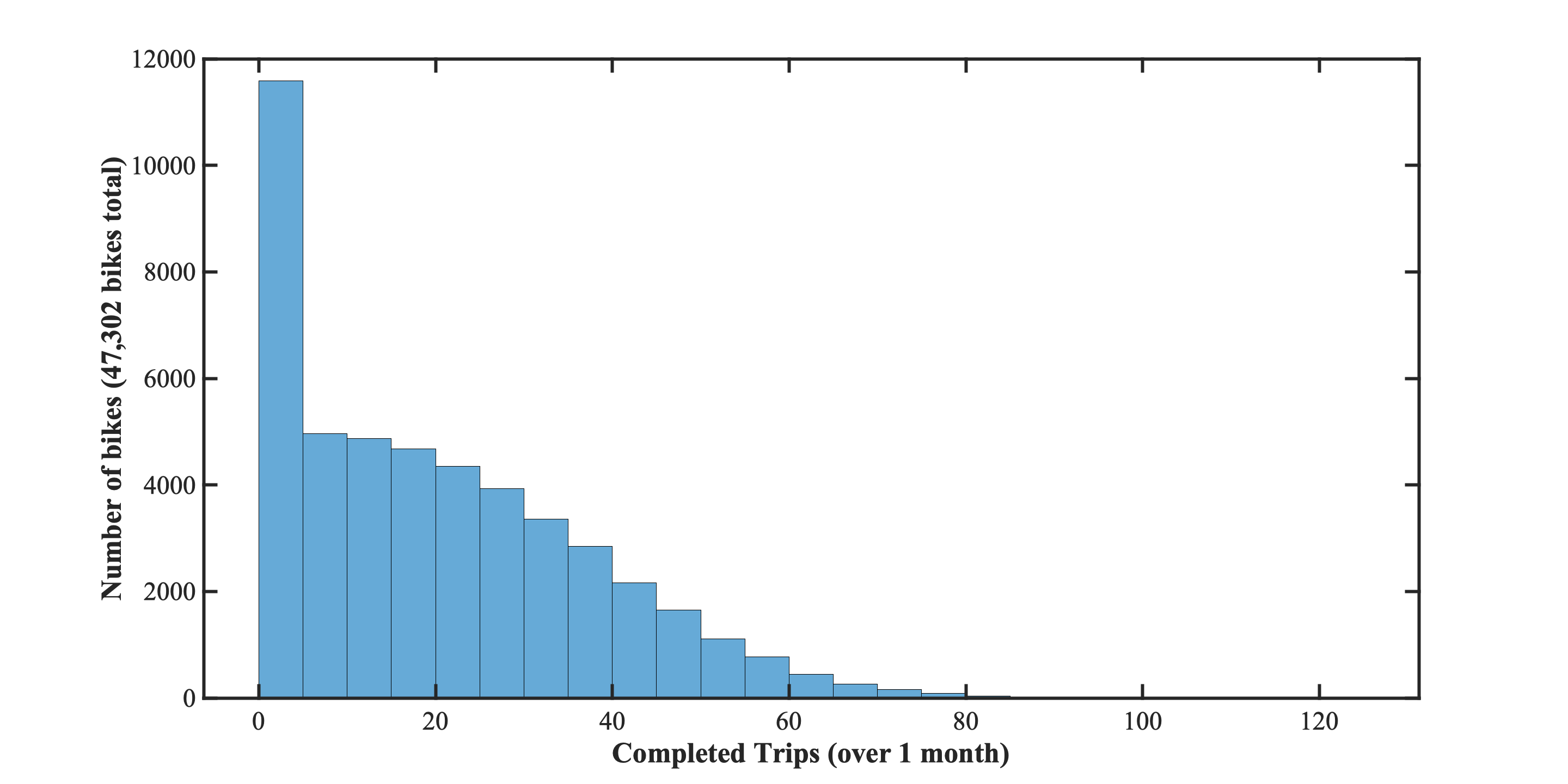}
\caption{Usage Frequency of bikes over a month}
\label{frequency}
\end{figure}

When considering the data for all of Singapore over the month of September 2017, we notice that usage of the 47,302 bikes was quite variable. Figure \ref{frequency} breaks down the large bike fleet in terms of the number of trips bikes were utilized for. While the most used bike was used 123 times that month, 11,575 (24.5\%) bikes were used $\leq 4$ times and 5,858 of those bikes were unused for the entire month. Possible explanations for the skewed distribution of Figure \ref{frequency} could be that: bikes broke and were left on the street; bikes were dropped in inaccessible locations (such as hidden behind a bush); there was an excess of bikes such that many of them did not need to get used. 

\begin{figure}[!h]
\centering
\includegraphics[width=.6\linewidth]{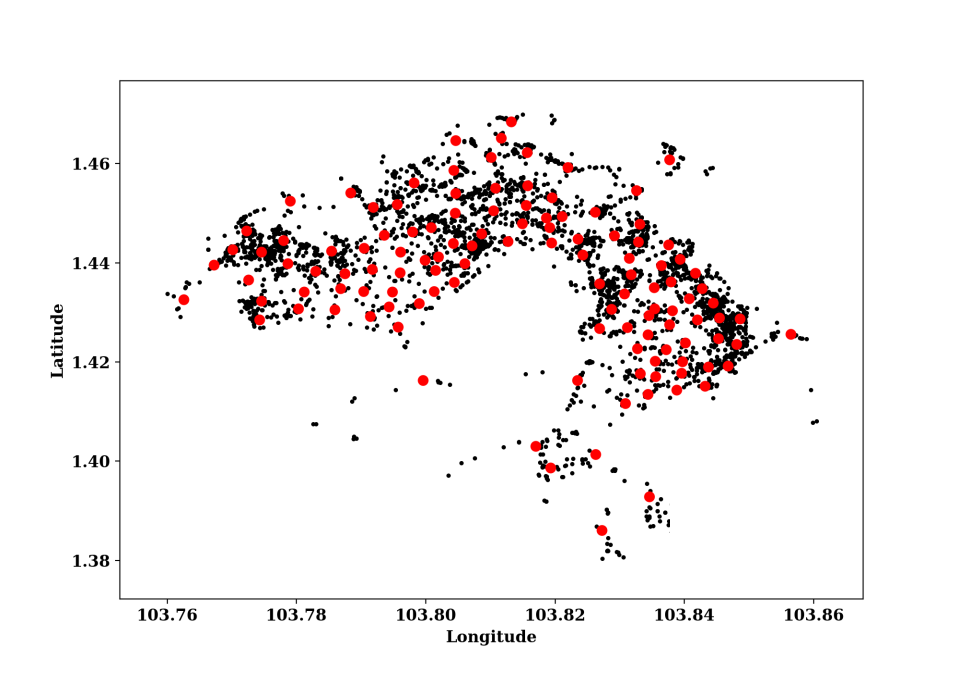}
\caption{Idle Bikes (black) and Abstract Stations (red)}
\label{idlebikes}
\end{figure}

To gain a better understanding of why bikes are unused in the Woodlands, we consider bikes idle for 7 or more consecutive days over the month of September and visualize their location alongside the abstract stations. As seen in Figure \ref{idlebikes}, a wide majority of the idle bikes were located at or near stations. It seems bikes were left in populated areas and remained unused as opposed to being left in distant areas and unused due to inaccessibility. Approximately 66\% of all bikes were idle for at least 7 days at some point during September 2017, and it seems unrealistic to propose that such a high portion of bikes are broken. We conclude that there was an excess of supply and infer that our data-set reflects the complete demand.

These trends in bike usage, as well as the guidelines of the LTA's operator license, served as motivation for investigating the impact of diminishing fleet size on the amount of lost demand in Singapore's dockless BSS. 

\section{Results}
\label{results}

For this section of the paper, we demonstrate the applicability of our MIP by taking the approach of a service-oriented provider in Singapore. As history shows (and data confirms), competitive Dockless BSS providers overflowed the city with bikes, leading to harsh regulations from the LTA. Hence, we estimate best-case-scenario lost demand for two rebalancing strategies at different fleet sizes. Furthermore, we compare the performance of dynamic and static rebalancing while reducing the fleet size in a step-wise manner. We modify the initial conditions by multiplying the number of bikes at each station by a set constant (ranging between 0.4 and 1) and flooring the result to deal with an integer amount of bikes.

Tables \ref{table_statdyn} and \ref{table_statdyn_constants} specify the setting configurations used to generate the results below. Table \ref{table_statdyn} deals with setting differentiation between static and dynamic repositioning in the MIP, while Table \ref{table_statdyn_constants} specifies the setting configurations kept the same between the two strategies. We attribute 15 vehicles to static repositioning to avoid setting an upper bound of repositioning trips taken while keeping a manageable problem size for the MIP. We consider the effects of changing the number of vehicles for the dynamic case later in this paper, but select 3 as a relatively realistic value for this section. For both strategies, the MIP is iterated 720 times with a time-step of 1 hour. This equates to running the simulation for 30 days.

\begin{table}[h]
\centering
\begin{tabular}{ ||p{5.5cm}||p{2cm}|p{2cm}||  }
 \hline
 & Static & Dynamic\\
 \hline
 Number of vehicles $||V||$   & 15 &   3\\
 \hline
 Rebalancing freq. $T$ (in hrs) &   24 & 1\\
 \hline
\end{tabular}
\caption{MIP setting differences between Static and Dynamic Rebalancing}
\label{table_statdyn}
\end{table}

\begin{table}[h]
\centering
\begin{tabular}{ ||p{6.5cm}||p{4cm}||  }
 \hline
 & Static and Dynamic\\
 \hline
 Number of demand scenarios $||\mathcal{K}||$ & 5 \\ 
 \hline
 Vehicle capacity $C_v^*$ &   10\\
 \hline 
 Maximum pickup points $\mathcal{N}_p$   & 1 \\
 \hline
 Maximum drop-off points $\mathcal{N}_d$   & 1 \\
 \hline
 Number of iterations & 720 \\
 \hline
\end{tabular}
\caption{MIP setting constants between Static and Dynamic Rebalancing}
\label{table_statdyn_constants}
\end{table}

\begin{figure}[h]
\centering
\includegraphics[width=1\linewidth]{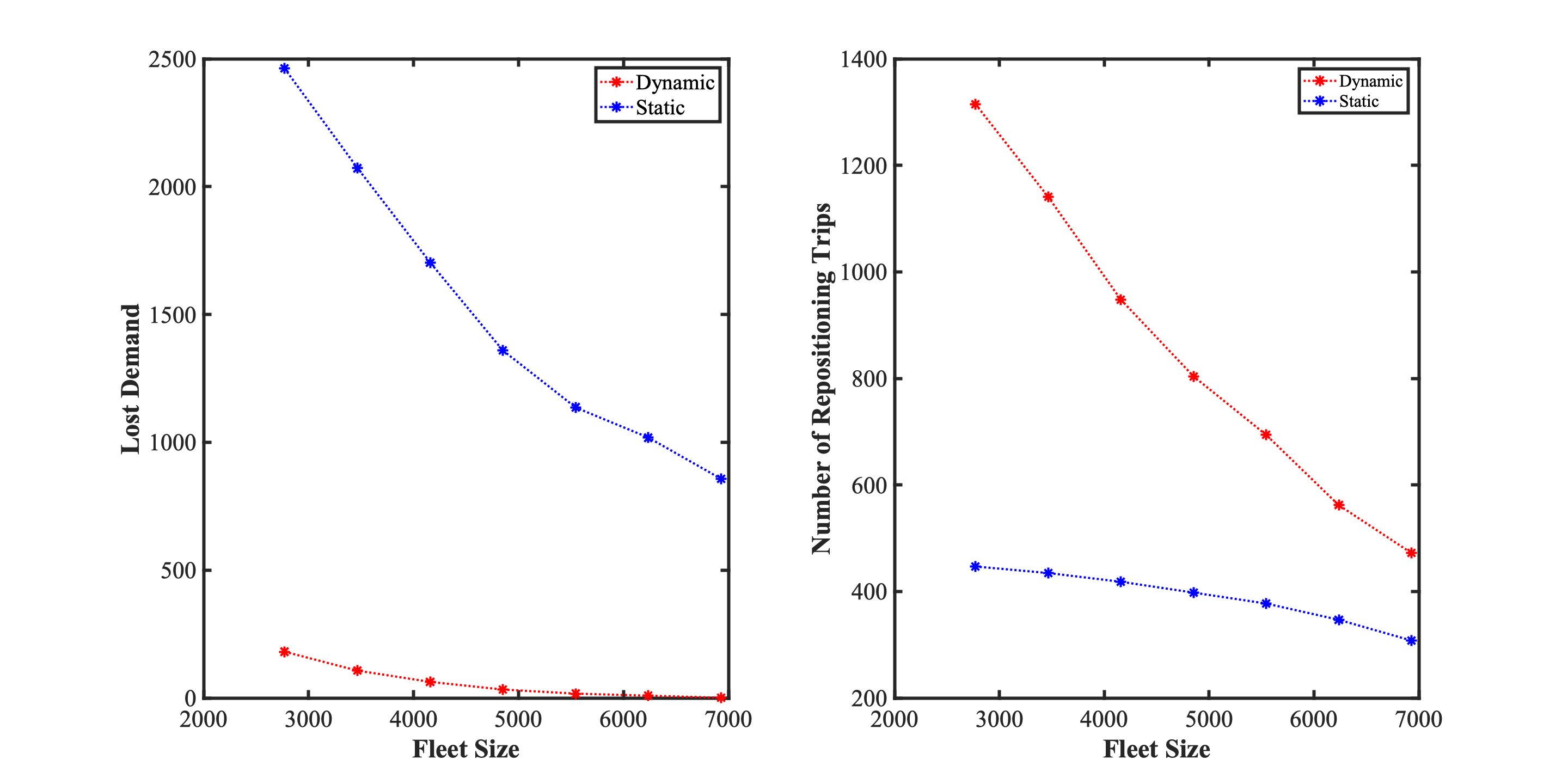}
\caption{Dynamic and Static Rebalancing under varying Fleet Size}
\label{vary_fleetsize_2plots}
\end{figure}

Figure \ref{vary_fleetsize_2plots} compares simulation results for static and dynamic rebalancing. Having a larger fleet is associated with a lower need of repositioning and lower lost demand. The relationship between fleet size and number of repositioning trips appears linear and negative in both cases ($R^2_{STATIC} = 0.966$ and $R^2_{DYNAMIC} = 0.986$), but the dynamic case is affected 6 times as much as the static one. The relationship between fleet size and lost demand is interesting as well. Both appear to be negative exponential ($R^2_{STATIC} = R^2_{DYNAMIC} = 0.966$), meaning the effects of increasing fleet size become less considerable as the fleet gets bigger.

\begin{figure}[t]
\centering
\includegraphics[width=1\linewidth]{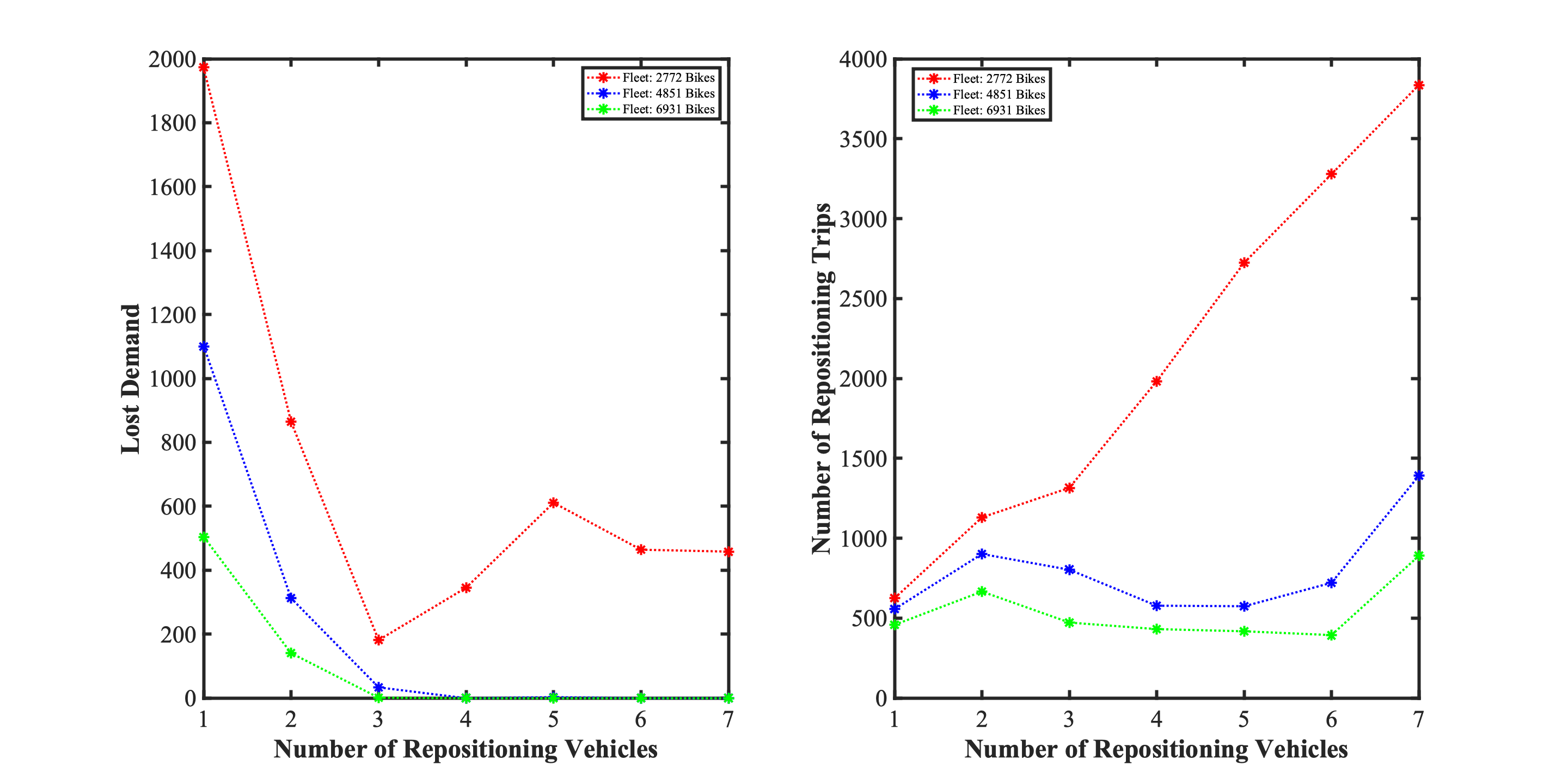}
\caption{Varying Bike and Vehicle Fleet Sizes}
\label{num_vehicles}
\end{figure}

Thinking about optimal fleet size, we also consider the effects of varying the number of repositioning vehicles ($||\mathcal{V}||$) involved in the dynamic repositioning. As displayed in Figure \ref{num_vehicles}, we consider fleet sizes of 6,931 (100\%), 4,851 (70\%) and 2,772 (40\%) and vary $||\mathcal{V}||$ from 1 to 7. For the two largest fleet sizes tested, the effects of changing number of vehicles are similar, but with a smaller fleet size of 2,772 bikes the effect of adding vehicles is more considerable and surprising. The red line (representing 2,772 bikes) in the Lost Demand plot of Figure \ref{num_vehicles} reaches a minimum when there are only 3 rebalancing vehicles in the system, while the number of repositioning trips taken increases linearly ($R^2 = 0.985$), insinuating that adding repositioning vehicles after a critical point worsens system performance. We infer that this trend is due to the repositioning frequency; in the dynamic scenario we study here, the MIP minimizes lost demand considering only the current hour. Increasing the repositioning potential might help minimize lost demand for some time-steps but destabilize the system and lead to unmanageable periods of lost demand at peak hours. Looking at the simulation data strengthens this hypothesis as peaks of large lost demand appear around peak hours. For larger bike fleet sizes, we infer that any vehicles added after 3 are essentially idle, as the lost demand approaches zero then and the number of rebalancing trips plateaus until $||\mathcal{V}|| = 6$. There is a spike in the number of rebalancing trips taken when $||\mathcal{V}|| = 7$, and we encourage further investigation in higher repositioning fleet sizes to understand this spike.  

\section{Future Work}
\label{future}
In this paper, we presented a viable framework for solving the repositioning problem for dockless BSS operators. Nevertheless, we see several areas to extend our work. The demand model used in our work assumed independence between stations, meaning the demand at stations is polled independently from geographically neighboring stations. This does not accurately model circumstances where there is a dependence in demand among stations. For example, if it is raining over a certain station and demand is low, it is very likely neighboring stations' demands will also suffer. Incorporating such correlated demand scenarios into $\mathcal{K}$ would yield more accurate rebalancing combinations. 

In addition, our definition of static repositioning is slightly different from the definition other related research, such as \cite{Ghosh}. Related research defines static repositioning as resetting the system to predefined states at night when demand is negligible. Optimizing BSS and analyzing the lost demand under such a definition requires a completely different linear program. We expect that the altered definition of static repositioning would perform better than our definition due to the higher granularity of where and how many bikes can get repositioned. However, the feasibility of such a system suffers greater than ours if many low-capacity (e.g. moving only 1 bike) repositioning trips are required to bring the system back exactly to the desired state.

To keep the problem simple, we made a deliberate decision to scale the initial station bike amounts linearly when generating the data-sets of varying fleet sizes. However, there are tendencies for repositioning trips to pick up bikes from a general sub-set of stations and drop-off at another sub-set. More intelligent fleet initial conditions could be generated than our linear model if these patterns are taken into consideration. Additionally, the 100\% fleet is entirely based on the initial placement of bikes in our data-set. It is likely sub-optimal and would require the same attention. Optimizing these initial conditions may produce better results. However, we decided to accept the simple linear transformation of initial station bike amounts since, over enough iterations of the MIP, the system will eventually converge to an optimal placement of bikes.

In addition, other clustering algorithms, such as community detection, could be used for defining abstract stations. The number of abstract stations used could also be increased to improve accuracy of the results.   

Thinking of Dockless BSS, an interesting research and optimization topic not discussed in this paper considers finding optimal routing strategies to pick-up/drop-off bikes, either statically or dynamically.

The linear program was designed to be as adaptable to different scenarios as possible and we believe it could be a useful tool for future research on dockless BSS and other dockless systems, such as electric scooters.

\section{Conclusion}

Dockless Bike Sharing Systems are an innovative, eco-friendly solution to transportation in urban environments. Rebalancing strategies have shown to improve the performance of docked BSS \cite{Ghosh}, and while the process may be more involved for dockless BSS, rebalancing can have a considerable impact in those systems as well. The mixed integer program and data preprocessing techniques we propose form a flexible framework to model dockless BSS trends and rebalancing strategies. The adjustable objective parameters allow for researchers and BSS providers to optimize for their own priorities whether it be customer satisfaction, profit or some mixture of both. Results from running the MIP model over data collected in the Woodlands region of Singapore revealed interesting relationships which may drive crucial operations decisions of BSS providers. For example, increasing the number of vehicles available to reposition bikes did not always improve the lost demand of the system for certain fleet sizes which may be counter-intuitive at first. In addition, adjusting the fleet size had a larger impact on the amount of rebalancing trips required to sustain the system for dynamic repositioning when compared to static repositioning. Overall, rebalancing strategies vary based on the environment and goals of the operator, and we hope our simulation framework will be a useful tool in comparing potential BSS approaches or facilitate related further research. 

The code for the data preprocessing, demand model and mixed integer program used to extract the simulation results in this paper is available upon request.

\section*{Acknowledgements}

We thank Professor Patrick Jaillet for the guidance and insightful discussions while working on this research. We also thank the Future Urban Mobility (FM) Interdisciplinary Research Group for hosting us during our research under the support of the National Research Foundation (NRF), Singapore under the Campus for Research Excellence and Technological Enterprise (CREATE) programme, Singapore-MIT Alliance for Research and Technology (SMART) Center. This research was organized as a part of the Singapore-MIT Undergraduate Research Fellowship (SMURF) program.

\end{document}